\documentclass[10pt]{article}
\usepackage[letterpaper]{geometry}
\usepackage{hicss}
\usepackage{times}
\usepackage[none]{hyphenat}
\usepackage{url}
\usepackage{latexsym}
\usepackage{minted}
\usepackage{indentfirst}
\usepackage{graphicx}
\graphicspath{{images/}}
\usepackage{amsmath,amssymb,amsfonts}
\usepackage{graphicx}
\usepackage{booktabs}
\usepackage{algorithm}
\usepackage{algorithmicx}
\usepackage{algpseudocode}
\usepackage{xcolor}
\usepackage{url}
\usepackage{cite}
\usepackage{array}
\usepackage{makecell}
\usepackage{dblfloatfix}

\usepackage{natbib}

\AtBeginDocument{
  \setlength{\abovedisplayskip}{3pt plus 1pt minus 1pt}
  \setlength{\belowdisplayskip}{3pt plus 1pt minus 1pt}
  \setlength{\abovedisplayshortskip}{2pt plus 1pt minus 1pt}
  \setlength{\belowdisplayshortskip}{2pt plus 1pt minus 1pt}
  \setlength{\jot}{1pt}
}

\title{Quantifying and Attributing Power Flexibility\\from GPU-Heavy Data Centers}

\author{
Yiru Ji \\
Georgia Institute of Technology \\
{\underline{yji332@gatech.edu}}
\And
Constance Crozier \\
Georgia Institute of Technology \\
{\underline{constance.crozier@gatech.edu}}
\And
Matthew Liska \\
Georgia Institute of Technology \\
{\underline{mliska@gatech.edu}}
}

\date{}

\begin{document}
\maketitle
\begin{abstract}
The growth of GPU-heavy data centers has  increased electricity demand and challenged grid stability. This paper investigates how an energy-aware job scheduling algorithm provides flexibility in GPU-heavy data centers. We develop a rolling-horizon optimization framework considering IT power and cooling dynamics with limited future job information. Compared with the first-in first-out baseline, we show that energy-aware scheduling brings latent power flexibility during peak-price periods. This flexibility is created through both thermal and computational mechanisms: cooling shifting can reliably reduce demand for short periods at relatively low incentive (\$30/MWh), and movement of backfilled jobs can often reduce demand at similar prices (\$30-300/MWh). Further reduction is possible through reordering or delaying jobs, but due to lost profits these actions come at higher prices (starting at \$600/MWh, more significantly above \$3000/MWh). Flexibility is achievable without knowing arriving jobs, but much greater flexibility can be achieved with perfect foresight of the future queue.
\end{abstract}
\vspace{1.2em}
\vspace{1.2em}
\noindent\textbf{Keywords:}  
Data center, energy optimization, demand response, job scheduling, grid flexibility.
\vspace{1.2em}
\raggedbottom
\section*{Nomenclature}
\small
\setlength{\tabcolsep}{3pt}
\renewcommand{\arraystretch}{1.08}

\noindent \textbf{Sets}\\[0.05cm]
\begin{tabular}{p{0.22\linewidth}p{0.70\linewidth}}
$\mathcal{J}$ & Set of all jobs\\
$\mathcal{J}_{\tau}$ & Set of jobs running at time $\tau$\\
$\mathcal{S}(j,\tau)$ & Set of start times for job $j$ be active at $\tau$\\
$\mathcal{T}$ & Set of all time steps\\
$\mathcal{T}^{\text{current}}$ & Set of time steps in the current window\\
$\mathcal{T}^{\text{peak}}$ & Set of time steps in the peak-price period\\
\end{tabular}

\vspace{0.25cm}
\noindent \textbf{Variables}\\[0.03cm]
\begin{tabular}{
>{\raggedright\arraybackslash}p{0.28\linewidth}
>{\raggedright\arraybackslash}p{0.64\linewidth}
}
% $G_{[a,b]}$ & Node reduction over the window $[a,b]$\\
% $G_{\text{peak}}$ & Peak-period node-usage reduction\\
% $G_{\text{recovered}}$ & Portion of peak-period node reduction recovered within the visible window\\
% $I_{\tau}$ & Average IT power intensity per occupied node at time $\tau$\\
% $N_{\tau}$ & Number of occupied nodes at time $\tau$\\
$p^{\text{cont}}_{j,\tau}$ & Power of continuing job $j$ at time $\tau$\\
$p^{\text{cooling}}_{\tau}$ & Thermal cooling power at time $\tau$\\
$p^{\text{IT}}_{\tau}$ & IT power consumption at time $\tau$\\
$p^{\text{new}}_{j,\tau}$ & Power of newly started job $j$ at time $\tau$\\
% \end{tabular}

% \vspace{0.08cm}
% \noindent
% \begin{tabular}{
% >{\raggedright\arraybackslash}p{0.28\linewidth}
% >{\raggedright\arraybackslash}p{0.64\linewidth}
% }
% $s^{\text{backfill}}$ & Share of positive node-occupancy saving attributed to backfilling\\
% $s^{\text{delay}}$ & Share of positive node-occupancy saving attributed to job delaying\\
$T^{\text{Tank}}_{\tau}$ & Tank temperature at time $\tau$\\
$t^{\text{start}}_j$ & Start time of job $j$\\
$x^{\text{cont}}_{j,\tau}$ & 1 if job $j$ continues from a previous window at time $\tau$, otherwise 0\\
$x^{\text{sched}}_{j}$ & 1 if job $j$ has already been scheduled, otherwise 0\\
$x^{\text{start}}_{j,\tau}$ & 1 if job $j$ starts at time $\tau$, otherwise 0\\
% $\Delta p^{\text{backfill}}$ & Power reduction from backfilling\\
% $\Delta p^{\text{cooling}}$ & Power reduction from cooling shifting\\
% $\Delta p^{\text{delay}}$ & Power reduction from job delaying\\
% $\Delta p^{\text{occ}}$ & Power reduction from node occupancy\\
% $\Delta p^{\text{occ-offset}}$ & Offset caused by higher occupancy\\
% $\Delta p^{\text{reordering}}$ & Power reduction from job reordering\\
% $\Delta p^{\text{reordering-offset}}$ & Offset caused by higher power intensity\\
\end{tabular}

\vspace{0.35cm}
\noindent \textbf{Parameters}\\[0.05cm]
\begin{tabular}{p{0.22\linewidth}p{0.70\linewidth}}
% $a$ & Start time of the visible analysis window\\
% $A$ & Heat exchange area\\
% $b$ & End time of the visible analysis window\\
% $C$ & Specific heat capacity of the coolant\\
$c_{\text{elec}}$ & Electricity price at time $\tau$ (\$/MWh)\\
$c_{\text{GPU}}$ & Revenue per GPU-hour (\$/GPU-hour)\\
% $COP$ & Coefficient of performance of cooling\\
$d_j$ & Duration of job $j$ (h)\\
$\Delta t$ & Length of one simulation time step (h)\\
% \end{tabular}

% \vspace{0.08cm}
% \noindent
% \begin{tabular}{
% >{\raggedright\arraybackslash}p{0.28\linewidth}
% >{\raggedright\arraybackslash}p{0.64\linewidth}
% }
$G_{\text{node}}$ & Number of GPUs per node\\
$g_j$ & Number of GPUs requested by job $j$\\
% $H$ & Rolling optimization window length\\
$N$ & Total number of available nodes\\
$n_j$ & Number of nodes required by job $j$\\
$p_{\text{GPU}}$ & Rated dynamic power per GPU (W)\\
$p_{\text{idle}}$ & Idle power consumption per node (W)\\
$t^{\text{end}}$ & End time of the current rolling window (h)\\
$T^{\text{Out}}_{\tau}$ & Outside temperature at time $\tau$ ($^\circ$C)\\
$t^{\text{start}}$ & Start time of the current rolling window (h)\\
$T^{\text{Tank,min/max}}$ & Minimum/maximum tank temperature ($^\circ$C)\\
$t^{\text{arr}}_j$ & Arrival time of job $j$ (h)\\
$t^{\text{wait}}$ & Maximum wait time allowed for jobs (h)\\
% $U$ & Heat transfer coefficient\\
$u_j$ & GPU utilization rate of job $j$\\
% $V^{\text{tank}}$ & Tank volume\\
% $\rho$ & Coolant density\\
\end{tabular}

\normalsize

\section{Introduction}

In recent years, data centers have emerged as critical infrastructure components for the digital economy. \citet{shehabi2024united} show that data center power consumption has grown rapidly with the expansion of artificial intelligence and cloud computing services. \citet{norris2025rethinking} indicate that data centers may consume approximately 10\% of the nationwide electricity demand by the end of this decade, creating growing challenges for grid operators. \citet{chen2025electricity} further emphasize that AI data centers differ from traditional data centers because their GPU-dense infrastructure creates high power density, making it harder to forecast demand and bringing potential challenges for grid planning and real-time stability.

\citet{wierman2014opportunities} demonstrate that data centers are well-suited for participation in demand response
programs since they are highly automated, and data centers continuously monitor and optimize the power loads, states of IT equipment, and cooling facilities. Unlike many industrial or commercial loads whose demand is tightly coupled to real-time physical activity, a portion of data center computation can be shifted in time \citep{lin2012dynamic}, across locations \citep{lin2012online}, or temporarily reduced without immediately interrupting end users \citep{he2012zeta}. Further, \citet{zhou2024ai} emphasize that AI-focused GPU-heavy data centers may be particularly well suited to provide flexibility because many AI workloads are naturally deferrable and queue-based. Given their substantial power consumption and various options for operational flexibility, \citet{takci2025data} conclude that data centers can act as flexibility assets for power systems through demand-side flexibility, energy flexibility, and ancillary-service participation.

Building on these opportunities, researchers have developed various approaches to incentivize and manage flexibility in data center operations. \citet{paul2015demand} examine incentive mechanisms for delay-tolerant workloads using an optimization framework, while \citet{guo2017colocation} develop a game-theoretic approach to balance operational costs against flexibility rewards while maintaining acceptable service quality. Other research considers contractual frameworks between data centers and grid operators \citep{basmadjian2016making}, leader-follower game models \citep{han2023two}, and learning-based approaches for dynamic resource management \citep{xu2018renewable}, all aimed at leveraging the inherent flexibility in data center operations.

Recent work has further explored the value of data center flexibility for power-system planning and operation. \citet{kim2026flexibility} show that standardized flexibility options, such as firm operation, pause, and shift, can expand data center siting opportunities and reduce interconnection constraints. \citet{wan2026data} further show that workload flexibility can help alleviate congestion, renewable curtailment, and operating costs.

However, much of the existing literature treats data center flexibility at an aggregate system level, representing flexible demand through simplified deferral windows, standardized load envelopes, or system-level shifting constraints. While these abstractions are useful for evaluating grid-level impacts, they provide limited insight into how flexibility is created internally through job-level scheduling decisions.

% To address these challenges, extensive research has explored the optimization of job scheduling within data centers to improve operational efficiency. Various algorithmic frameworks have been proposed, ranging from greedy heuristics \cite{dong2015greedy}, mixed-integer linear programming \cite{guo2025integrated}, to reinforcement learning that adapts to dynamic operating conditions \cite{shi2024energy, ran2019deepee, liu2023online}. Although traditional approaches primarily emphasized performance metrics such as throughput, latency, and resource utilization, recent work has evolved toward energy-aware scheduling strategies that consider electricity costs and carbon emissions. These strategies incorporate cooling control \cite{shi2024energy}, renewable energy utilization \cite{xu2018renewable}, and participation in electricity markets \cite{yuan2020revenue, lee2024online}, recognizing the complex interplay between computational workloads, energy systems, and external market conditions. This research shows that jobs with varying time sensitivity and resource requirements can create opportunities for flexible scheduling approaches that can intelligently prioritize and allocate resources \cite{yang2025pricing}.

\citet{crozier2025potential} identify several possible flexibility mechanisms, such as pre-cooling, altering CPU/GPU clock rate, and smart scheduling, but limited work has systematically quantified how different operational mechanisms contribute to the observed flexibility of GPU-heavy data centers. This paper addresses this gap by developing an energy-aware rolling-horizon scheduling framework for GPU-heavy data centers, considering the uncertainty in job arrivals, and examining the operational sources of their flexibility. Specifically, we:
\begin{enumerate}
    \item Quantify the flexibility potential of GPU-heavy data centers with queueable workloads.
    \item Develop a rubric which decomposes flexibility into the underlying operational mechanisms, including cooling shifting, job reordering, backfilling, and job delaying, and quantify their contributions across different price incentives.
    \item Analyze the sensitivity of this potential to key operational parameters in data centers, such as queue length and job heterogeneity.
\end{enumerate}

\section{Baseline data center operation}

This section introduces the basic components in data centers and scheduling logic that is currently most commonly used. This paper focuses on data centers that compute large queueable jobs on GPUs that don't need to start instantly, such as AI training or HPC simulations.

\subsection{Parameter definitions}

%A GPU-heavy data center operates as a complex system, where understanding key operational concepts is essential for optimizing resource allocation and energy efficiency.

Nodes and GPUs represent the fundamental computing architecture in a data center. A node typically refers to a complete physical server, including processors, memory, local storage, and networking components. GPUs are specialized accelerators installed within these nodes. A typical high-performance computing node may contain one to eight GPUs, interconnected via high-speed buses. Workloads usually request resources based on the number of GPUs. When a job cannot fully utilize all the GPUs in the allocated node, resource fragmentation occurs, reducing overall efficiency.

Node occupancy reflects the proportion of available nodes currently running jobs, expressed as the ratio of allocated nodes to the total number of nodes. High occupancy indicates efficient use of the physical infrastructure, but may limit the system's flexibility in responding to new job scheduling requests.

GPU utilization reflects the extent to which GPU power is actually being utilized, measuring how close the GPU is operating to its thermal design power limit. For example, computationally intensive tasks high in matrix arithmetic and floating point operations can often max out the thermal design power, leading to high GPU power utilization. In contrast, operations like memory transfers, even at peak memory bandwidth, typically consume less power and thus result in lower power utilization.

The cooling system is responsible for managing the large amount of heat generated by the computing hardware. As the GPU density and power intensity increase, so do the cooling requirements. Modern data centers employ a variety of strategies, but as traditional air cooling methods become inadequate, operators typically switch to liquid cooling and in the future also immersion cooling.

\subsection{First-in first-out scheduling}

The first-in first-out (FIFO) with backfilling algorithm is currently widely used in high-performance computing job scheduling. In basic FIFO, jobs are processed strictly in arrival order, which might lead to resource under-utilization when large jobs block smaller ones. Backfilling addresses this limitation by allowing smaller or lower-priority jobs to use idle resources ahead of time, but only if doing so does not delay the expected start of any earlier arriving jobs. This is achieved by calculating a timeline based on the estimated completion times of running jobs. If a gap appears before a high-priority job needs its resources, a qualifying lower-priority job can be scheduled into this gap immediately \citep{jette2023architecture}.

\section{Energy-aware scheduling algorithm}

In this section, we introduce our proposed energy-aware scheduling algorithm. First, we formulate an optimization model to determine optimal job start times within a static time window. Second, we describe the rolling-horizon procedure that applies this model sequentially across the window.

\subsection{Optimization over a static time window}

Here we assume the scheduler knows all job information, including: job arrival times, GPU utilization, node requirements, and durations.
% This includes both currently available real jobs and anticipated future jobs, which are modeled as a proxy for future workload pressure.
Further, we also consider that some jobs may already be running from the previous window. For each current time window $\mathcal{T}^{\text{current}}$, we aim to find the job execution schedule that will maximize the following objective:

\begin{align*}
\max Z =& \sum_{\substack{j \in \mathcal{J}: x^{\text{sched}}_j=0}} c_{\text{GPU}} \cdot g_j \cdot d_j \cdot \sum_{\tau \in \mathcal{T}^{\text{current}}} x^{\text{start}}_{j,\tau} \\
&- \sum_{\tau \in \mathcal{T}^{\text{current}}}  c_{\text{Elec}} \cdot \left(p^{\text{IT}}_{\tau} + \frac{p^{\text{cooling}}_{\tau}}{COP} \right)
\end{align*}

This function reflects profit by considering job revenue, electricity, and cooling costs. The revenue term calculates the total revenue from scheduling jobs, where $x^{\text{sched}}_{j}$ is a binary variable indicating whether job $j$ has been scheduled in a previous time window or not (1 if scheduled, 0 otherwise), $c_{\text{GPU}}$ represents the monetary value per GPU-hour, $g_j$ denotes the number of GPUs requested by job $j$, $d_j$ is the duration of job $j$, and $x^{\text{start}}_{j,\tau}$ is a binary variable indicating whether job $j$ starts at time $\tau$. Therefore, the first term in the objective function represents the profit gained from jobs starting in the current time window $\mathcal{T}^{\text{current}}$, and the second term considers the electricity and cooling cost.

The IT power consumption at time $\tau$ is defined as:
\begin{align*}
p^{\text{IT}}_{\tau} &= p_{\text{idle}} \cdot N + \sum_{j \in \mathcal{J}} \left( p^{\text{cont}}_{j,\tau} + p^{\text{new}}_{j,\tau} \right)\\
\text{where,}\\
p^{\text{cont}}_{j,\tau} &= x^{\text{cont}}_{j,\tau} \cdot g_j \cdot u_j \cdot p_{\text{GPU}}\\
p^{\text{new}}_{j,\tau} &= g_j \cdot u_j \cdot p_{\text{GPU}} \cdot \sum_{s \in \mathcal{S}(j,\tau)} x^{\text{start}}_{j,s}\\
\mathcal{S}_{j,\tau}  &= \left\{ s \in \mathcal{T}^{\text{current}} : \max(t^{\text{arr}}_j, \tau-d_j+1) \leq s \leq \tau \right\}
\end{align*}

Where $p_{\text{idle}} \cdot N$ represents the idle power consumption of all available nodes, plus a dynamic component proportional to the number of actively utilized GPUs. For each job $j$ and time step $\tau$, the set $\mathcal{S}_{j,\tau}$ contains all possible start times $s$ within the current time window that would result in job $j$ being active at $\tau$. Specifically, a start time $s$ ensures the job starts no earlier than its arrival time and no later than $\tau$, while also ensuring the job's duration would cover the required $\tau$ time steps. Therefore, the power of active nodes can be calculated by both jobs continuing from previous time windows $p^{\text{cont}}_{j,\tau}$ and jobs starting within the current window $p^{\text{new}}_{j,\tau}$, where $u_j$ is the GPU utilization rate and $p_{\text{GPU}}$ is the rated dynamic power per GPU, which is calculated as the difference between maximum node power and idle node power divided by the number of GPUs per node.

The total cooling power is calculated by scaling the thermal cooling power $p^{\text{cooling}}_{\tau}$ by the coefficient of performance $COP$. This term measures how efficiently a cooling system uses electrical energy to move thermal energy. The electricity cost is calculated using a time-varying electricity price $c_{\text{elec}}$.

The scheduling problem is subject to the following operational constraints:
\begin{align}
&\sum_{\tau \in \mathcal{T}^{\text{current}}} 
x^{\text{start}}_{j,\tau}
\leq 1, 
\forall j \in \mathcal{J}: x^{\text{sched}}_j=0
\label{eq:c1}
\\
&x^{\text{start}}_{j,\tau} = 0,  
\quad 
\forall j \in \mathcal{J}, 
\forall \tau \in \mathcal{T}^{\text{current}}:
\tau < t^{\text{arr}}_j
\label{eq:c2}
\\
&\begin{aligned}
\sum_{\tau \in \mathcal{T}^{\text{current}}} 
\tau x^{\text{start}}_{j,\tau}
&\leq t^{\text{arr}}_j + t^{\text{wait}},\\
&\quad
\forall j \in \mathcal{J}: 
x^{\text{sched}}_j = 0,\ 
t^{\text{arr}}_j \leq t^{\text{end}}
\end{aligned}
\label{eq:c3}
\\
&\sum_{j \in \mathcal{J}} 
\left(
x^{\text{cont}}_{j,\tau}
+
\sum_{s \in \mathcal{S}_{j,\tau}} 
x^{\text{start}}_{j,s}
\right)n_j
\leq N,
\forall \tau \in \mathcal{T}^{\text{current}}
\label{eq:c4}
\\
%\end{align}
%\begin{align}
&\begin{aligned}
T^{\text{Tank}}_{\tau+1}
=&\ T^{\text{Tank}}_\tau
+ \frac{\Delta t}{\rho C V^{\text{tank}}}
\Big[
p_\tau^{\text{IT}}
- p^{\text{cooling}}_\tau \\
&\quad
+ U A
\left(
T^{\text{Out}}_\tau
-
T^{\text{Tank}}_\tau
\right)
\Big],
\forall
\tau \in \mathcal{T}^{\text{current}}
\end{aligned}
\label{eq:tank_dynamics}
\\
&T^{\text{Tank,min}}
\leq
T^{\text{Tank}}_{\tau+1}
\leq
T^{\text{Tank,max}},
\forall \tau \in \mathcal{T}^{\text{current}}
\label{eq:tank_limit}
\end{align}

Constraint \eqref{eq:c1} ensures that each unscheduled job can start at most once within the current time window. Constraint \eqref{eq:c2} prevents jobs from starting before their specific arrival times $t^{\text{arr}}_j$. Constraint \eqref{eq:c3} ensures that each job should start within the maximum waiting time $t^{\text{wait}}$ after its arrival. %This constraint only applies to jobs whose arrival time $t^{\text{arr}}_j$ is before the end of the current time window $T^{\text{end}}$.

The running status of jobs is tracked through the variable $x^{\text{cont}}_{j,\tau}$, which indicates whether job $j$ is active at time $\tau$, accounting for both jobs continuing from previous windows and jobs starting within the current window. Constraint \eqref{eq:c4} guarantees that the total nodes required by all running jobs does not exceed the total available node capacity $N$, where $n_j$ represents the number of nodes required by job $j$.

Constraint \eqref{eq:tank_dynamics} tracks the change of the cooling tank temperature. The variable $T^{\text{Tank}}_{\tau}$ represents the tank temperature at time $\tau$. The tank temperature increases with the generated heat load, which is proportional to IT power, and decreases according to the output of the cooling system. The final term accounts for heat exchange with the outside environment, where $T^{\text{Out}}_{\tau}$ is the outside temperature and $UA$ represents the overall thermal conductance. The term $\rho C V^{\text{tank}}$ represents the total thermal capacity of the tank, determined by the coolant density $\rho$, specific heat capacity $C$, and tank volume $V^{\text{tank}}$.  Constraint \eqref{eq:tank_limit} ensures that the tank temperature remains within its minimum allowable tank temperature $T^{\text{Tank,min}}$ and maximum allowable tank temperature $T^{\text{Tank,max}}$.

\subsection{Rolling horizon algorithm}

The rolling horizon algorithm is designed to solve large-scale job scheduling problems by decomposing the entire time horizon into smaller, manageable windows. The static-window model described above is used as the subproblem solved at each rolling step. This approach both allows us to solve for longer time horizons and can control the job queue dynamically as new jobs arrive over time.

Let $\mathcal{T}$ represent the complete set of all time steps in the scheduling horizon, and $H$ denote the rolling horizon length, which defines the size of each optimization window (e.g., $H=24$ time steps). The algorithm proceeds iteratively, solving an optimization problem for each time window while updating state information between consecutive windows.  

At each iteration $k = 0,1,2,\ldots,\lfloor (|\mathcal{T}|-1)/H \rfloor$, the current time window $\mathcal{T}^{\text{current}} = [t^{\text{start}}, t^{\text{end}}]$ is defined, where $t^{\text{start}} = k \cdot H$ and $t^{\text{end}} = \min((k+1)\cdot H - 1, |\mathcal{T}|)$. The algorithm maintains state variables including job scheduling status ($x^{\text{sched}}_{j}$) and start times ($t^{\text{start}}_j$). %Before solving each window's optimization, the algorithm updates the status of continuing jobs that started in previous windows but are still running in the current window. 

We distinguish between real jobs and anticipated future jobs in the rolling-horizon procedure. Real jobs represent workloads that have actually arrived and can be scheduled and executed by the operator. Anticipated future jobs are temporary proxy jobs generated within the look-ahead window to represent expected future workload pressure. 

Let $\mathcal{J}$ denote the full realized workload used in the simulation, including jobs with future arrival times that are not yet visible to the scheduler. At each rolling step, only the subset of jobs that has arrived is available for scheduling, denoted by $\mathcal{J}^{\text{real}}_{\text{arrived}}$. To account for future workload within the look-ahead window, the scheduler generates a temporary anticipated job set $\mathcal{J}^{\text{future}}$. The optimization is solved using both $\mathcal{J}^{\text{real}}_{\text{arrived}}$ and $\mathcal{J}^{\text{future}}$, but only decisions for real jobs are executed. The anticipated future jobs are used only to guide current scheduling decisions and are discarded after each rolling step.

The details of the algorithm are described below.
\begin{algorithm}
\caption{Rolling horizon algorithm}
\begin{algorithmic}[1]
\State \textbf{Initialize} $x^{\text{sched}}_{j} = 0$, $t^{\text{start}}_j = -1$ $\forall j \in \mathcal{J}$
\For{each time window $k = 0,1,2,\ldots,\lfloor (|\mathcal{T}|-1)/H \rfloor$}
    \State $[t^{\text{start}}, t^{\text{end}}] = [k \cdot H, \min((k+1)\cdot H - 1, |\mathcal{T}|)]$
    \State $\mathcal{T}^{\text{current}} = \{t^{\text{start}}, \dots, t^{\text{end}}\}$
    \State \textbf{Update} $x^{\text{cont}}_{j,\tau}=1$ if $x^{\text{sched}}_{j}=1$ and $t^{\text{start}}_j < t^{\text{start}} \leq \tau < t^{\text{start}}_j + d_j$, else $0$, $\forall j \in \mathcal{J}$
    \State \textbf{Identify} arrived real jobs $\mathcal{J}^{\text{real}}_{\text{arrived}}$
    \State \textbf{Generate} anticipated future jobs $\mathcal{J}^{\text{future}}$
   
    \State \textbf{Solve single window optimization model} to decide $x^{\text{start}}_{j,\tau}$ for 
$\{j \in \mathcal{J}^{\text{real}}_{\text{arrived}} \mid x^{\text{sched}}_j = 0\} \cup \mathcal{J}^{\text{future}}$

    \For{each real job $j \in \mathcal{J}^{\text{real}}_{\text{arrived}}$ where $\sum_{\tau \in \mathcal{T}^{\text{current}}} x^{\text{start}}_{j,\tau} = 1$}
        \State \textbf{Update} $x^{\text{sched}}_{j} = 1$ and $t^{\text{start}}_j = \tau$ where $x^{\text{start}}_{j,\tau} = 1$
    \EndFor
    \State \textbf{Discard} $\mathcal{J}^{\text{future}}$
\EndFor
\end{algorithmic}
\end{algorithm}

% \begin{algorithm}
% \caption{Rolling Horizon Algorithm}
% \begin{algorithmic}[1]
% \State \textbf{Initialize} $x^{\text{sched}}_{j} = 0$, $t^{\text{start}}_j = -1$ $\forall j \in \mathcal{J}$
% \For{each time window $k = 0,1,2,\ldots,\lfloor (|\mathcal{T}|-1)/H \rfloor$}
%     \State $[T^{\text{start}}, T^{\text{end}}] = [k \cdot H, \min((k+1)\cdot H - 1, |\mathcal{T}|)]$
%     \State $\mathcal{T}^{\text{current}} = \{T^{\text{start}}, \dots, T^{\text{end}}\}$
%     \State \textbf{Update} $x^{\text{cont}}_{j,\tau}=1$ if $x^{\text{sched}}_{j}=1$ and $t^{\text{start}}_j < T^{\text{start}} \leq \tau < t^{\text{start}}_j + d_j$, else $0$
%     \State \textbf{Solve single window optimization model} to determine $x^{\text{start}}_{j,\tau}$ for $\{j \in \mathcal{J} \mid x^{\text{sched}}_j = 0\}$
%     \For{each job $j$ where $\sum_{\tau \in \mathcal{T}^{\text{current}}} x^{\text{start}}_{j,\tau} = 1$}
%         \State \textbf{Update} $x^{\text{sched}}_{j} = 1$ and $t^{\text{start}}_j = \tau$ where $x^{\text{start}}_{j,\tau} = 1$
%     \EndFor
% \EndFor
% \end{algorithmic}
% \end{algorithm}

In brief, each iteration of the rolling horizon algorithm performs three tasks: (1) identify which jobs are continuing from previous windows, (2) solve the optimization in the current time window to schedule new job starts, and (3) update the status of all jobs.

\section{Simulation framework} 

In this section, we describe how the simulation was set up, the parameters that we chose in our experiments, and the method we used to evaluate flexibility.

\subsection{Simulation setup}

Our simulations focus on a GPU-heavy data center with queueable loads operating over multiple days with a rolling horizon optimization approach. The simulation process begins with generating a realistic workload of heterogeneous jobs with varying resource requirements, durations, and arrival times.

Table \ref{tab:simulation_params} summarizes the key parameters used in our simulation environment. We synthesized a workload of 150 jobs with arrival times distributed across a 5-day time window. The time step granularity of our simulation is 1 hour, and the total simulation length is 120 hours. To better represent the limited future information, the scheduler uses a 10-hour rolling optimization window and rolls forward every 1 hour. At each rolling step, we add an anticipated future job queue to represent uncertainty and approximate potential future arrivals within the look-ahead window.

The data center is equipped with 100 nodes with 4 GPUs per node. The maximum and idle power of each node is 3000W \citep{nvidiah200} and 900W respectively \citep{shehabi20242024}. The coolant-related data is sourced from the cooling system of the Frontier supercomputer \citep{28e9db15b4684466a91ec1c65e2b415c}.

Job characteristics follow distributions derived from analysis of real-world GPU cluster workloads. We implemented an inverse relationship between the number of GPUs requested and the probability of job occurrence: jobs requesting fewer GPUs appear more frequently than those demanding larger GPU amounts. Furthermore, we assume that 20\% of jobs will terminate early before their estimated completion time.

The anticipated future job queue is used only to inform the rolling algorithm about the likely future workload pressure. Otherwise, the scheduler will systematically underestimate future workloads. As the anticipated jobs always have future arrival times, they will never be executed; they only help the scheduler avoid overly myopic decisions when future arrivals are uncertain. 

% \begin{table}[ht]
% \centering
% \caption{Simulation Parameters}
% \label{tab:simulation_params}
% \begin{tabular}{lc}
% \hline
% \textbf{Parameter} & \textbf{Value} \\
% \hline
% \multicolumn{2}{l}{\textit{Rolling Window}} \\
% Total simulation length & 120 hours \\
% Rolling horizon window & 24 hours \\
% Time step granularity & 1 hour \\
% \hline
% \multicolumn{2}{l}{\textit{Data Center Parameters}} \\
% Number of compute nodes & 100 \\
% GPUs per node & 4 \\
% Maximum power per node & 3000W \\
% Idle power per node & 900W \\
% Cooling coefficient & 0.4 \\
% \hline
% \multicolumn{2}{l}{\textit{Job Parameters}} \\
% Mean job duration & 10 hours \\
% Standard deviation of duration & 6 hours \\
% Maximum job duration & 48 hours \\
% Maximum wait time & 30 hours \\
% GPU utilization mean & 0.6 \\
% Percentage of early terminating jobs & 20\% \\
% Job pricing & \$2.30 / GPU-hour \\
% \hline
% \end{tabular}
% \end{table}

\begin{table}[!t]
\centering
\caption{Parameter setup for the simulation.}
\label{tab:simulation_params}
\begin{tabular}{lc}
\hline
\textbf{Parameter} & \textbf{Value} \\
\hline
\multicolumn{2}{l}{\textit{Rolling Window}} \\
Total simulation length & 120 hours \\
Rolling optimization window & 10 hours \\
Rolling step length & 1 hour \\
Time step granularity & 1 hour \\
\hline
\multicolumn{2}{l}{\textit{Data Center Parameters}} \\
Number of compute nodes & 100 \\
GPUs per node & 4 \\
Maximum power per node & 3000 W \\
Idle power per node & 900 W \\
\hline
\multicolumn{2}{l}{\textit{Thermal Parameters}} \\
Tank volume & 10 $\text{m}^3$ \\
% Initial tank temperature & $30^{\circ}\text{C}$ \\
Minimum tank temperature & $12^{\circ}\text{C}$
 \\
Maximum tank temperature & $27^{\circ}\text{C}$
 \\
Outside temperature & $20^{\circ}\text{C}$
 \\
Coefficient of performance & 20 \\
Coolant density & 1060 $\text{kg}\cdot\text{m}^{-3}$ \\
Coolant specific heat capacity & 3500 $\text{J}/(\text{kg} \cdot \text{K})$ \\
Thermal conductance & 40 $\text{W/K} $\\
\hline
\multicolumn{2}{l}{\textit{Real Job Parameters}} \\
Number of real jobs & 150 \\
Mean job duration & 10 hours \\
Standard deviation of duration & 6 hours \\
Maximum job duration & 48 hours \\
Maximum wait time & 30 hours \\
GPU utilization mean & 0.6 \\
Percentage of early terminating jobs & 20\% \\
Job pricing & \$2.30 / GPU-hour \\
\hline
\multicolumn{2}{l}{\textit{Anticipated Future Workload Parameters}} \\
Arrival rate & 1 job/hour \\
Duration & 10 hours \\
GPU requirement & 50 \\
GPU utilization & 0.6 \\
Maximum wait time & 30 hours \\
\hline
\end{tabular}
\end{table}

All experiments were conducted on a system equipped with an ROG Zephyrus G16 GU603VV laptop using Python 3.8 with Gurobi 11.0.3 as the solver.

\subsection{Testing flexibility}

To evaluate the flexibility of the data center, we simulate a sudden peak price scenario, which can occur in modern power grids when demand surges, generation fluctuates, or transmission lines become constrained. Data centers with flexible workload management capabilities could reduce demand during these events.

In our experiment, we introduce a significant peak price in the middle of our simulation time range, where the electricity price triples from the baseline rate of \$10/MWh to \$30/MWh for a duration of 1 hour. This short-lived price increase simulates a grid-stress event, where the timing of the event is known, as in day-ahead pricing.

We quantify the system's flexibility by measuring the difference between the power usage during the peak price period under the dynamic pricing scenario and the power usage during the same hour in a flat-price scenario. This difference represents the system's ability to temporarily reduce load in response to price signals. We also examine how the overall operation of the data center is affected outside of the peak period.

\subsection{Flexibility attribution mechanisms}
\begin{figure}[]
\centering
\includegraphics[width=1\linewidth]{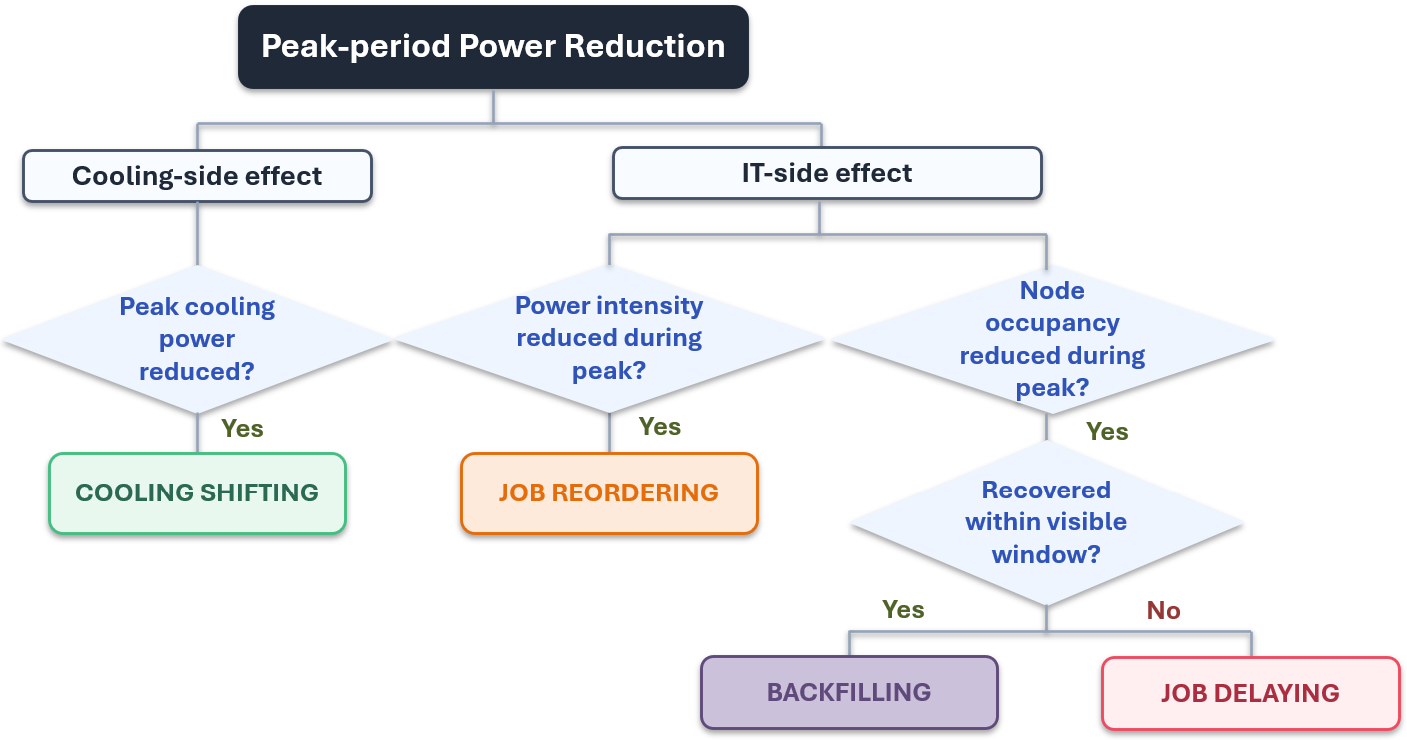}
\caption{Flexibility attribution mechanisms.}
\label{fig:mechanism_flow}
\end{figure}

After measuring the total power reduction during the peak price period, we further decompose where this flexibility comes from, as shown in Figure~\ref{fig:mechanism_flow}. In our framework, the reduction can come from four operational mechanisms: cooling shifting, job reordering, backfilling, and job delaying. We calculate their numerical contributions based on the peak-period power reduction. 
% We report these mechanisms as signed contributions. A positive value indicates that the mechanism reduces peak-period power, while a negative value indicates that it offsets reductions created by other mechanisms.

\subsubsection{Cooling shifting}

Cooling shifting refers to the reduction in cooling power during the peak price period. In our model, this flexibility can come from the thermal storage effect. The system can pre-cool before the price spike, allowing the tank temperature to rise during the peak period while using less cooling power. Therefore, the contribution from cooling shifting is viewed as the reduction in cooling power during the peak period, which can be measured as the difference between the cooling power under flat price $p^{\text{cooling, flat}}_{\tau}$ and cooling power under peak price $p^{\text{cooling, peak}}_{\tau}$:
\begin{align*}
\Delta p^{\text{cooling}}
=
\frac{1}{|\mathcal{T}^{\text{peak}}|}
\sum_{\tau \in \mathcal{T}^{\text{peak}}}
\frac{
\left(
p^{\text{cooling, flat}}_{\tau}
-
p^{\text{cooling, peak}}_{\tau}
\right)}{COP}
\end{align*}

\subsubsection{Job reordering}

Instead of reducing the number of occupied nodes, the scheduler may choose jobs with a lower rate of GPU utilization during peak-price hours, thus lowering power consumption. In this case, the data center is still running jobs, but  consumes less power per occupied node. To quantify this effect, we write the dynamic IT power as: $p^{\text{IT}}_{\tau}=N_{\tau} I_{\tau}$ where $N_{\tau}$ is the number of occupied nodes and $I_{\tau}$ is the average IT power intensity per occupied node. Specifically:
\begin{align*}
N_{\tau}
&=
\sum_{j \in \mathcal{J}_{\tau}}
\frac{g_j}{G_{\text{node}}},I_{\tau}
=
\frac{p^{\text{IT}}_{\tau}}{N_{\tau}}\
\end{align*}
note that when $N_{\tau}=0$, we set $I_{\tau}=0$.

The job reordering contribution $\Delta p^{\text{reordering}}$ is then observed as the reduction in power while keeping the peak-price node occupancy level unchanged:
\[
\Delta{p}^{\text{reordering}}
=
\frac{1}{|\mathcal{T}^{\text{peak}}|}
\sum_{\tau \in \mathcal{T}^{\text{peak}}}
N^{\text{peak}}_{\tau}
\left(
I^{\text{flat}}_{\tau}
-
I^{\text{peak}}_{\tau}
\right)
\]
A positive value means that the algorithm runs lower-intensity jobs during the peak period. A negative value means that the peak-price case has higher GPU utilization than the flat-price one during the peak period.

% If $\Delta \tilde{p}^{\text{reordering}}<0$, the peak-price case has higher GPU utilization than the flat-price case during the peak period. In this situation, the reordering effect does not contribute to power reduction. Instead, it offsets part of the reduction from other mechanisms, so we report it as an reordering offset:
% \[
% \Delta p^{\text{reordering-offset}}
% =
% \min
% \left(
% \Delta \tilde{p}^{\text{reordering}},0
% \right)
% \]

\subsubsection{Mobile backfilling gaps}

When small jobs are backfilled, there may be latency in the job start time (i.e. the job could be shifted without delaying any future jobs). In this case where the scheduler reduces node occupancy during the peak period and recovers the reduced node occupancy later within the visible analysis window. Node occupancy change $\Delta p^{\text{occ}}$ is calculated as:
\[
\Delta {p}^{\text{occ}}
=
\frac{1}{|\mathcal{T}^{\text{peak}}|}
\sum_{\tau \in \mathcal{T}^{\text{peak}}}
I^{\text{flat}}_{\tau}
\left(
N^{\text{flat}}_{\tau}
-
N^{\text{peak}}_{\tau}
\right)
\]
% \[
% \Delta p^{\text{occ}}
% =
% \max \left(0, \Delta \tilde{p}^{\text{occ}}
% \right)
% \]

% If $\Delta \tilde{p}^{\text{occ}}>0$, the peak-price case uses fewer nodes during the peak period. 

To distinguish whether the reduced peak-period node usage is recovered later, we define the net node reduction over a generic period $\mathcal{K}$ as: \[ G_{\mathcal{K}} = \sum_{\tau \in \mathcal{K}} \left( N^{\text{flat}}_{\tau} - N^{\text{peak}}_{\tau} \right). \]

We then separate this node occupancy effect into backfilling and delaying. Let $\mathcal{T}^{\text{peak}}$ denote the peak price window and $[a,b]$ denote the visible analysis window. In our experiments, we remove the first and last day of simulated operation from our analysis to avoid boundary effects. 
% The peak-period node-usage reduction $G_{\text{peak}}$ is:
% \begin{align*}
% G_{\text{peak}}
% =
% \sum_{\tau \in \mathcal{T}^{\text{peak}}}
% \max
% \left(
% N^{\text{flat}}_{\tau}
% -
% N^{\text{peak}}_{\tau},
% 0
% \right)\,,
% \end{align*}
% and the node reduction over the window $G_{[a,b]}$ is:
% \begin{align*}
% G_{[a,b]}
% =
% \sum_{\tau=a}^{b}
% \left(
% N^{\text{flat}}_{\tau}
% -
% N^{\text{peak}}_{\tau}
% \right)
% \end{align*}
The recovered portion of nodes during the visible window is $
\max
\left(
G_{\text{peak}}
-
\max(G_{[a,b]},0),
0
\right)$.
Therefore, the backfilling share $s^{\text{backfill}}$ is:
\begin{align*}
s^{\text{backfill}}
=
\begin{cases}
\min \left( \frac{\max
\left(
G_{\text{peak}}
-
\max(G_{[a,b]},0),
0
\right)}{G_{\text{peak}}}, 1 \right),  G_{\text{peak}}>0\\
0
, G_{\text{peak}}=0
\end{cases}
\end{align*}
and the backfilling contribution $ p^{\text{backfill}}
=
s^{\text{backfill}}
\Delta p^{\text{occ}}$.

\subsubsection{Job delaying}

Job delaying captures the part of the peak-period node occupancy reduction that is not recovered within the visible window. This means that the scheduler reduces power during the peak by postponing some workload beyond the analysis window.

The delaying share is the inverse of the backfilling share, $s^{\text{delay}}=1-s^{\text{backfill}}$ and the delaying contribution is $\Delta p^{\text{delay}}
=
s^{\text{delay}}
\Delta p^{\text{occ}}$.

% If $\Delta \tilde{p}^{\text{occ}}<0$, the peak-price case occupies more nodes than the flat-price case during the peak period. In this situation, the node occupancy effect does not contribute to power reduction. Instead, it offsets part of the reduction from other mechanisms, so we report it as an occupancy offset:
% \begin{align*}
% \Delta p^{\text{occ-offset}}
% =
% \min
% \left(
% \Delta \tilde{p}^{\text{occ}},0
% \right)
% \end{align*}

% Combining all components, the total peak-period power reduction is attributed as:
% \begin{align*}
% \Delta p^{\text{total}}
% =
% \Delta p^{\text{cooling}}
% +
% \Delta p^{\text{reordering}}
% +
% \Delta p^{\text{backfill}}
% +
% \Delta p^{\text{delay}}
% +
% \Delta p^{\text{reordering-offset}}
% +
% \Delta p^{\text{occ-offset}}
% \end{align*}

% This attribution allows us to distinguish whether the observed flexibility mainly comes from thermal storage, lower-power job selection, schedule rearrangement, or workload postponement.

\section{Results}
In this section, we present the simulation results, attribute the observed flexibility to different mechanisms, and examine how information and workload conditions affect flexibility.

% We compare our scheduling algorithm against the FIFO with backfilling model under both the flat and peaking electricity price signals, using the data center simulation described above.

% Our analysis focuses on multiple metrics including average power consumption, average GPU utilization rate, average node occupancy rates, and total revenue. We examined both the complete simulation duration (excluding the first and last time window to mitigate boundary effects) and the peak price period for targeted analysis.

\subsection{Overall performance comparison}

% \textcolor{blue}{CC: I think we are lacking a time-series figure. In an ideal world perhaps a time-series of power consumption for the middle 24 hours Opt vs FIFO highlighting the peak period. A simpler version of one of the ones we had before.}

\begin{figure}[H]
\centering
\includegraphics[width=0.9\linewidth]{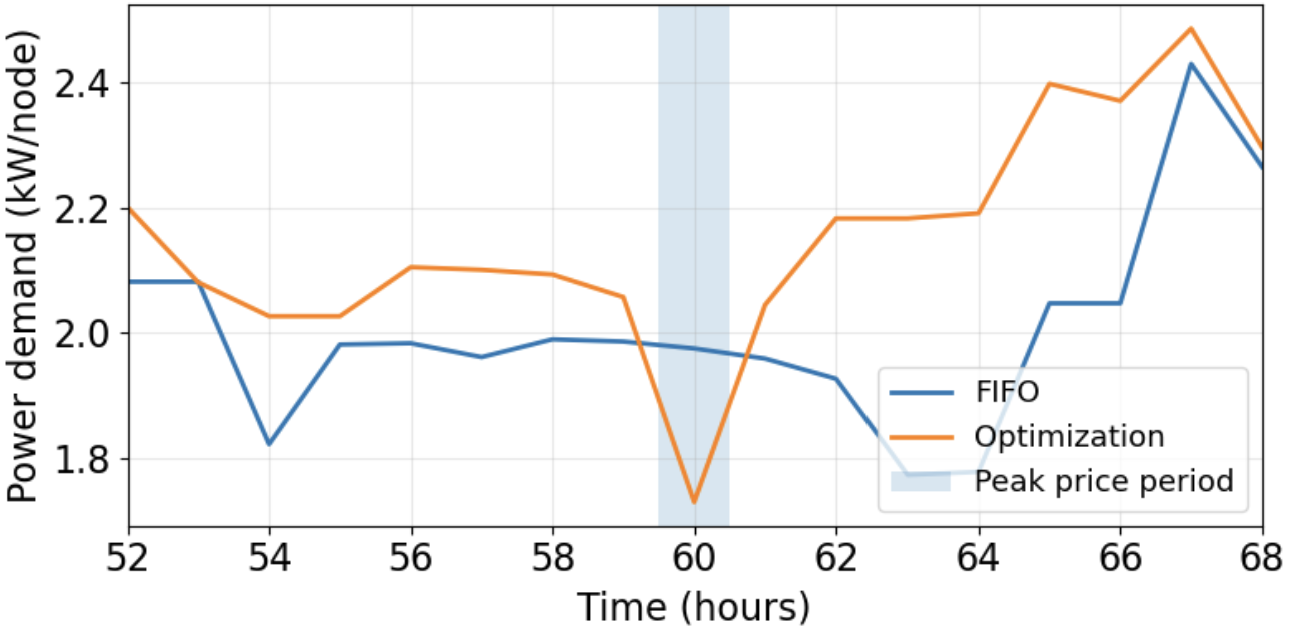}
\caption{Power demand around the peak-price period under FIFO and energy-aware optimization.}
\label{fig:power_demand}
\end{figure}

We compare our scheduling algorithm against FIFO with backfilling, using the data center simulation described above. Figure~\ref{fig:power_demand} shows the change in power demand between the FIFO and the energy-aware optimization method around the peak price period. The optimization model reduces demand during the peak hour and shifts part of the workload to surrounding non-peak periods.

We further compare the methods in Table \ref{tab:period_comparison}. Metrics are presented for both the complete simulation duration (excluding the first and last time window to mitigate boundary effects) and the peak price period for targeted analysis. Across the whole simulation period, we observe that both methods achieve similar levels of average power usage, GPU utilization, and node occupancy. Our optimization method reduces power during the peak period (5\% reduction), at the cost of longer waiting times and a very small drop in total revenue (0.1\%). It should be noted that, as discussed later, the size of the peak reduction is dependent on price and job characteristics.

\begin{table}[H]
\centering
\caption{Performance comparison between the whole period and peak price period.}
\label{tab:period_comparison}
\small
\begin{tabular}{l|cccc}
\hline
& \multicolumn{2}{c}{All Period} & \multicolumn{2}{c}{Peak Price Period} \\
\cline{2-5}
& Opt. & FIFO & Opt. & FIFO \\
\hline
Avg power (kW/node) & 1.88 & 1.88 & 1.90 & 2.00 \\
GPU utilization & 0.59 & 0.59 & 0.62 & 0.60 \\
Node Occupancy & 0.78 & 0.78 & 0.83 & 0.88 \\
Avg Wait (h) & 3.21 & 1.36 & - & - \\
Median Wait (h) & 1.59 & 1.13 & - & - \\
Max Wait (h) & 22.7 & 7.10 & - & - \\
Revenue (k\$) &52.36 & 52.42 & 0.77 & 0.90\\
\hline
\end{tabular}
\end{table}
%During the peak price period, the optimization method has lower power consumption, achieved mainly by decreasing the overall node occupancy when the electricity price is high.

% \begin{table}[htbp]
% \centering
% \caption{Performance Comparison between the Whole Period and Peak Price Period}
% \label{tab:period_comparison}
% \small
% \begin{tabular}{l|cccc}
% \hline
% & \multicolumn{2}{c}{All Period} & \multicolumn{2}{c}{Peak Price Period} \\
% \cline{2-5}
% & Opt. & FIFO & Opt. & FIFO \\
% \hline
% Average Power (kW) & 176 & 175 & 153 & 164 \\
% GPU Utilization & 0.55 & 0.55 & 0.52 & 0.51 \\
% Node Occupancy & 0.74 & 0.74 & 0.56 & 0.69 \\
% Average Wait (h) & 1.95 & 1.67 & - & - \\
% Revenue (k\$) & 49.87 & 49.32 & 0.51 & 0.64 \\
% \hline
% \end{tabular}
% \end{table}

\subsection{Mechanism attribution analysis}

\begin{figure}[H]
\centering
\includegraphics[width=1\linewidth]{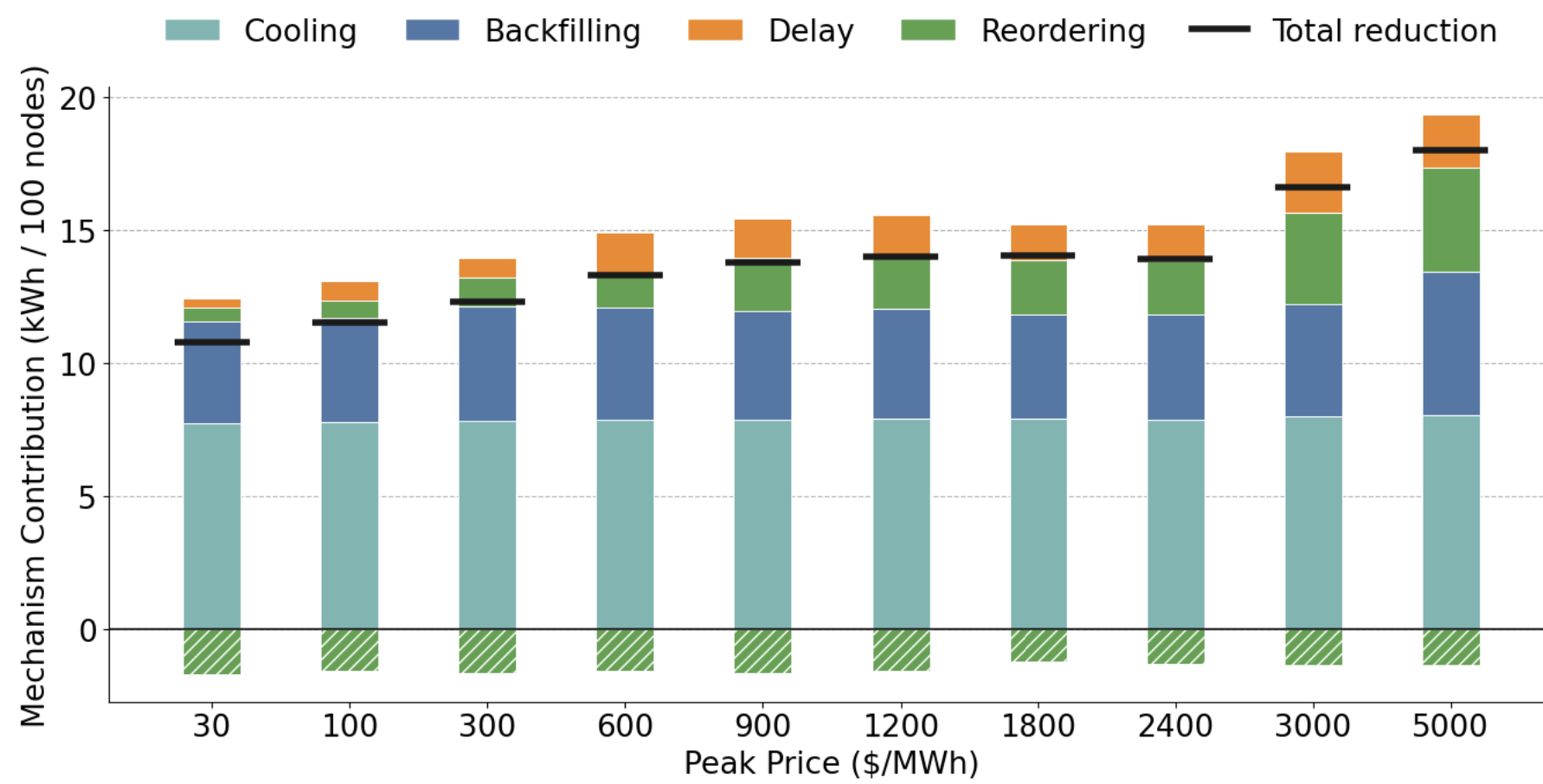}
\caption{Mechanism contribution to peak-period power reduction under different peak prices.}
\label{fig:mechanism_attribution}
\end{figure}

To understand where the observed peak-period power reduction comes from, we decompose the total reduction into cooling shifting, backfilling, job delaying, job reordering, and the corresponding offset terms according to the heuristic method presented in Seciton 4. Figure \ref{fig:mechanism_attribution} shows the mechanism contribution decomposition across different peak prices. The results show that the total power reduction during the peak period is achieved through a combination of both thermal and computational scheduling mechanisms, rather than by a single dominant source. 

Cooling shifting provides the most stable contribution across different peak prices. It reduces power by 8kW per 100 nodes at the price of \$30/MWh and no further increase is achieved at higher prices. This indicates that cooling system is less sensitive to price event and provides a relatively consistent baseline source of flexibility through thermal storage.

Backfilling also contributes significantly and is relatively stable across most price levels. The backfilling contribution remains around 4 kW when price changes between \$30/MWh to \$3000/MWh, and increases to 5.13 kW when the peak price reaches \$5000/MWh. This suggests that the amount of recoverable node shifting is more constrained by job timing and capacity gaps than the price multiplier. Backfilled jobs have shorter duration and use a smaller number of GPUs compared to the average job. Backfilling potential is therefore likely dictated by the number of such jobs in the queue. 

We observe negative values for job reordering across all prices. There are likely many equivalent solutions to the optimal queue under our current framework, and these values likely reflect attribution artifacts arising when the job order changes without affecting the peak power demand. Some positive contributions from reordering and delaying may also be partly offset by these negative values. Therefore, we believe that at prices below \$600/MWh we primarily see flexibility from backfilling and cooling. This range of prices is comparable to generator pricing, so flexibility may be routinely provided via these mechanisms. 

Between \$600 and \$2400/MWh we see a small amount of additional flexibility through reordering of the queue and/or slight delays in jobs. However, these mechanisms only begin to increase significantly at prices of \$3000/MWh and higher. These prices are typically higher than generator costs, although still below the market cap in most systems. Therefore, these mechanisms are likely to be useful in extreme cases of congestion or demand shortfall.

Job reordering likely leads to introduced latency (unless there happens to be a 1:1 swap of two jobs whose only difference is in GPU utilization) and so both mechanisms are likely trading off against the job revenue. The exact price at which this flexibility is deployed will depend heavily on the assumed revenue per GPU-hour; here we considered a relatively high value consistent with a commercial machine, however in-house private AI-training machines (whose profit isn't directly linked to jobs) may provide these mechanisms at a lower price point. 

%Furthermore, backfilled jobs tend to be shorter and lower-value than the average job. Their average duration is 15.35\% lower, their average GPU requirement is 5.85\% higher, and their average job value is 26.27\% lower than the overall average. This suggests that the scheduler tends to backfill jobs that are shorter and easier to fit into the remaining visible window, even though they require slightly more GPUs. 

%As the peak price increases, job-level scheduling mechanisms become more important. The reordering contribution increases from 0.49 kW at \$30/MWh to 3.96 kW at \$5000/MWh. This trend shows that stronger price signals encourage the scheduler to preferentially run jobs with lower power intensity during peak period. 

%Job delaying also becomes more prominent under higher peak prices. The delaying contribution increases from 0.34 kW at \$30/MWh to 2.71 kW at \$5000/MWh. This indicates that when the price signal is sufficiently strong, the scheduler increasingly reduces peak-period power by postponing some workload beyond the visible analysis window. Therefore, flexibility at higher peak prices may come at the cost of reduced service quality and longer job waiting times.

%The reordering offset is negative across all peak price levels. This means that after the scheduler reduces node occupancy during the peak period, the remaining jobs may have a higher average power intensity than those in the flat-price case. This higher intensity partially offsets the power reduction created by running fewer nodes.

\subsection{Impact of job uncertainty}

Next we examine how information availability affects the data center's flexibility potential. The previous experiments use a realistic online setting, where the scheduler rolls forward every hour with a 10-hour optimization window, and only currently arrived real jobs are known. We compare this setting with two perfect-information cases, where all real jobs are known in advance, but with one having a larger planning window. 

\begin{table}[htbp]
\centering
\caption{Flexibility under different levels of job information and rolling horizon settings.}
\label{tab:info_horizon_flexibility}
\small
\renewcommand{\arraystretch}{1.25}
\begin{tabular}{
@{}
>{\centering\arraybackslash}m{0.23\linewidth}
>{\centering\arraybackslash}m{0.25\linewidth}
>{\centering\arraybackslash}m{0.30\linewidth}
>{\centering\arraybackslash}m{0.10\linewidth}
@{}
}
\hline
\textbf{Case} 
& \textbf{Job Certainty} 
& \textbf{Setting} 
& \textbf{Flex.} \\
\hline

Online rolling 
& Imperfect info
& \makecell[c]{10-hour window\\1-hour step}
& 12.1 \\

Daily rolling 
& Perfect info 
& \makecell[c]{24-hour window\\5 steps}
& 26.2 \\

Full-horizon
& Perfect info
& \makecell[c]{120-hour window\\1 step}
& 71.3 \\

\hline
\end{tabular}
\vspace{0.2em}
\begin{flushleft}
\footnotesize Note: Flexibility is reported in kWh per 100 nodes.
\end{flushleft}
\end{table}

As shown in Table~\ref{tab:info_horizon_flexibility}, the realistic online setting provides the lowest flexibility, with a reduction of 12.1 kWh per 100 nodes. %This reflects a conservative estimate under limited future information, since the scheduler only observes arrived jobs and approximates future workloads using an anticipated future job queue.
When perfect job information is available, the estimated flexibility increases substantially. %With a 24-hour rolling horizon and 5 rolling steps, the flexibility increases to 26.2 kWh because the scheduler can better determine which jobs should be advanced, postponed, or reordered during the peak price period. 
This is especially true when the scheduler has access to the full window of arriving jobs in advance, rather than each day at a time. This demonstrates that the realistically deployable flexibility reported in the previous section is actually only a fraction of the true flexibility potential. 

For data centers operating multi-day queues and with a reduced number of users whose jobs are predictable, these higher levels of flexibility may be achievable by including prediction algorithms. Large scientific HPC facilities and in-house AI training centers may be good candidates for these, since they typically have a smaller number of users submitting homogeneous jobs. Commercial data centers which execute jobs from many different users may have a more limited potential to provide flexibility. 

\begin{table*}[!t]
\centering
\caption{Comparison during peak period with varied GPU utilization variance, number of jobs, and GPUs per job.}
\label{tab:peak}
\small
\setlength{\tabcolsep}{2.5pt}
\begin{tabular}{@{}c|ccc|ccc|ccc|ccc|ccc|ccc@{}}
\toprule
\textbf{} &
\multicolumn{3}{c|}{\textbf{Util=0.6}} &
\multicolumn{3}{c|}{\textbf{Util$\sim$N(0.6,0.3)}} &
\multicolumn{3}{c|}{\textbf{Jobs=100}} &
\multicolumn{3}{c|}{\textbf{Jobs=200}} &
\multicolumn{3}{c|}{\textbf{GPUs/job=20}} &
\multicolumn{3}{c}{\textbf{GPUs/job$\sim$P(20)}} \\
&
FIFO & \multicolumn{2}{c|}{Opt} &
FIFO & \multicolumn{2}{c|}{Opt} &
FIFO & \multicolumn{2}{c|}{Opt} &
FIFO & \multicolumn{2}{c|}{Opt} &
FIFO & \multicolumn{2}{c|}{Opt} &
FIFO & \multicolumn{2}{c}{Opt} \\
{\color[HTML]{9B9B9B} Peak Price} &
{\color[HTML]{9B9B9B}} & {\color[HTML]{9B9B9B} 3} & {\color[HTML]{9B9B9B} 300} &
{\color[HTML]{9B9B9B}} & {\color[HTML]{9B9B9B} 3} & {\color[HTML]{9B9B9B} 300} &
{\color[HTML]{9B9B9B}} & {\color[HTML]{9B9B9B} 3} & {\color[HTML]{9B9B9B} 300} &
{\color[HTML]{9B9B9B}} & {\color[HTML]{9B9B9B} 3} & {\color[HTML]{9B9B9B} 300} &
{\color[HTML]{9B9B9B}} & {\color[HTML]{9B9B9B} 3} & {\color[HTML]{9B9B9B} 300} &
{\color[HTML]{9B9B9B}} & {\color[HTML]{9B9B9B} 3} & {\color[HTML]{9B9B9B} 300} \\
\midrule
Power (kW) &
169 & 156 & 138 &
164 & 153 & 125 &
151 & 121 & 109 &
199 & 189 & 138 &
182 & 147 & 124 &
169 & 135 & 119 \\
GPU Util &
0.55 & 0.55 & 0.55 &
0.51 & 0.52 & 0.42 &
0.62 & 0.57 & 0.52 &
0.59 & 0.57 & 0.45 &
0.65 & 0.66 & 0.65 &
0.57 & 0.58 & 0.49 \\
Occupancy &
0.69 & 0.57 & 0.41 &
0.69 & 0.57 & 0.36 &
0.49 & 0.27 & 0.17 &
0.88 & 0.82 & 0.49 &
0.67 & 0.43 & 0.28 &
0.67 & 0.39 & 0.28 \\
Revenue (\$) &
646 & 516 & 363 &
646 & 516 & 311 &
511 & 223 & 144 &
805 & 749 & 433 &
660 & 398 & 255 &
641 & 335 & 258 \\
\bottomrule
\end{tabular}
\end{table*}

\subsection{Sensitivity to job characteristics}

To examine how workload characteristics affect data center flexibility, we conduct a sensitivity analysis under the perfect-information setting, where the scheduler optimizes over
five 24-hour rolling windows. Table \ref{tab:peak} compares the impact of three job factors: GPU utilization variance, job queue length, and the number of GPUs requested per job.

\subsubsection{GPU utilization variance}

We compare the case where all jobs have the same utilization to the case where utilization follows a normal distribution with the same mean. This shows that heterogeneity in job GPU utilization is a key driver for flexibility. A workload composed of jobs with diverse GPU utilization levels allows the scheduler to selectively execute low-utilization jobs during peak price periods, thereby enabling a 23.8\% reduction in power consumption without interrupting service. By contrast, if all jobs have the same utilization, then only 18.3\% reduction can be created from having nodes idle during the peak period. 

% As shown in Figure \ref{fig:spike_GPU_sensitive}, for moderate price increases, such as a price multiplier between 3 and 30, changes in GPU utilization variance have minimal impact on power consumption patterns. However, when the peak price multiplier reaches 300, a workload with high GPU utilization variance enables a 23.8\% reduction in power consumption, significantly outperforming the 18.3\% reduction achieved under a homogeneous utilization profile. This demonstrates that in extreme peak price scenarios, systems with higher GPU utilization variance offer greater opportunities for power optimization.

% \begin{figure}[htbp]
% \centering
% \includegraphics[width=0.9\linewidth]{spike_GPU_sensitive_1222.png}
% \caption{Percentage change of average power during the peak price period compared with FIFO under different peak prices and GPU variance.}
% \label{fig:spike_GPU_sensitive}
% \vspace{-0.3cm}
% \end{figure}

\subsubsection{Job queue length}

We also test the effect of job queue length, comparing cases with 100 or 200 jobs total instead of the baseline 150. When the job queue length is shorter, the system can provide more flexibility. With 100 jobs, the queue is shorter and the data center is less fully occupied, our method reduces power consumption by 19.6\% compared to FIFO. When the total job number is 200 and the data center is highly occupied, the room for flexibility is reduced and the power reduction is only 4.9\%. The less heavily used a data center is, the more latent flexibility can be provided.

\subsubsection{Relative job size}

Job-size heterogeneity, measured by the number of GPUs requested per job, also affects peak-period power. Under a 3x price multiplier, the optimization achieves a 19.2\% power reduction for fixed-size jobs (20 GPUs each) and a 20.1\% reduction for variable-size jobs (Poisson-distributed with mean 20). The corresponding peak-period power decreases from 147 kW to 135 kW, suggesting that variation in GPU demand can give the scheduler more opportunities to adjust job placement during the peak period.

%Under an extreme 300x multiplier, the reductions are 31.9\% and 29.6\%, respectively. 
% This indicates that job-size diversity is a less significant driver of flexibility compared to the others we study in the analysis. The algorithm can effectively shift or defer jobs regardless of their GPU count, provided there is sufficient slack in the queue. 
% Thus, while job-size variability may slightly alter the mix of jobs that run during the peak period, it does not substantially enhance the data center's flexibility.

\section{Conclusion}

In this paper, we demonstrate that energy-aware scheduling can enhance grid flexibility of GPU-heavy data centers with queueable workloads. Compared with the baseline FIFO method, our algorithm reduces power consumption during peak price periods, at the cost of a longer waiting time and a slight reduction in revenue, while maintaining similar overall utilization. This shows that GPU-heavy data centers can provide latent flexibility through operational scheduling decisions, even when the future job information is limited.

We further show that the observed flexibility is created through multiple mechanisms. At relatively low incentive levels, the dominant flexibility sources are cooling shifting and backfilling. Cooling shifting provides a stable contribution by using the thermal storage during the peak period at a very low incentive (\$30/MWh). Backfilling also makes a significant and solid contribution at similar price levels by rearranging shorter and smaller jobs. The two mechanisms can provide power reductions at prices comparable to generator costs, suggesting that they may be available for routine grid-support events. Additional flexibility from job reordering and delaying accounts for a small proportion, appears around \$600/MWh, and increases substantially only at higher prices above \$3000/MWh in our experiments, indicating that these mechanisms are more likely to be used during more extreme conditions.

Our results also highlight the importance of information availability. The realistic online rolling setting shows that flexibility can be gained without knowing future arrivals. Perfect-information optimization produces substantially higher flexibility, indicating that more flexibility can be achieved with a better prediction of future job. Through sensitivity analysis, we identify two key factors that increase flexibility potential: (1) data centers with lower queue utilization, which provide greater scheduling freedom, and (2) workloads with higher variance in job characteristics, particularly
GPU utilization rates. 

These results suggest that the grid value of data center flexibility depends on different aspects, including the price signal, workload composition and future job information. However, current results only use synthetic workloads. Future work can validate the findings with real GPU cluster traces to bridge the gap between simulated benchmarks and empirical operations.

\addtolength{\textheight}{-.2cm} 

%Bibliography 
\bibliographystyle{apalike}
\bibliography{sample}

% % \printbibliography

% \section{References}

% \addtolength{\textheight}{-.2cm}

% \printbibliography[heading=none]

\end{document}